# Interpreted Programming Language Extension for 3D Render on the Web

A. Duarte and E. Ramírez

*Abstract*—There are tools to ease the 2D/3D graphics development for programmers. Sometimes, these are not directly accessible for all users requiring commercial licenses or based on trials, or long learning periods before to use them. In the modern world, the time to release final programs is crucial for the company successfully, also for saving money. Then, if programmers can handle tools to minimize the development time using well-known programming languages, they can deliver final programs on time, with minimum effort. This concept is the goal of this paper, offering a tool to create 3D renders over a familiarize programming language to speed up the web development time process. We present an extension of an interpreted programming language with an easy syntax to display 3D graphics on the web generating a template in a well-known web programming language, which can be customized and extended. Our proposal is based on Lua programming language as the input language for programmers, offering a web editor which interprets its syntax and exporting templates in WebGL over Javascript, also getting immediate output in a web browser. Tests show the effectiveness of our approach focus on the written code lines, also getting the expected output using a few computational resources.

## I. INTRODUCCIÓN

PARA un desarrollador de aplicaciones gráficas 3D, el proceso de *rendering*, que va desde la creación de las estructuras de datos hasta conseguir una visualización en algún dispositivo de salida, no es una tarea trivial y resulta extensa en el número de líneas de código. Esto ha creado una necesidad por construir herramientas que faciliten y reduzcan la cantidad de líneas de código. Existen diferentes herramientas que cumplen ambos o uno de estos puntos, ya sea abstrayendo al desarrollador del código o, proporcionándole un código sencillo base para construir otros más complejos.

Existen diversos lenguajes de programación y extensiones de lenguajes que soportan el despliegue de gráficos. Las extensiones de lenguajes consisten en expansiones de otro lenguaje de programación, ofreciendo un ambiente de desarrollo conocido por los programadores, siendo ampliamente disponibles y fáciles de mantener. Sin embargo, estas extensiones suelen ser complejas de utilizar y comprender, además de extender los códigos en demasiadas líneas. En contraparte, los lenguajes enfocados solo en el despliegue de gráficos 3D (lenguajes completos), suelen ser más expresivos en el dominio gráfico y más simples su desarrollo; pero suelen poseer licencias comerciales, siendo no accesibles para cualquier usuario.

Ahora, si ambas opciones reducen la cantidad de líneas de código, también acarrean inconvenientes. Las extensiones sólo pueden ser empleadas a través del lenguaje que extienden, acarreando los inconvenientes del lenguaje. El lenguaje puede ser desconocido por el programador, de lógica complicada y no permitir visualizar los resultados de la ejecución de manera inmediata. Los lenguajes nuevos para el programador lo obligan a aprender este nuevo lenguaje, pudiendo no proveer las funcionalidades que requiere en su despliegue. Igualmente, pudiera no proveer una forma de exportar el despliegue a otro lenguaje o herramienta conocida donde el desarrollador si pueda implementar o utilizar la funcionalidad que requiere.

Los APIs de despliegue gráfico (e.g. OpenGL, DirectX) generalmente se basan un lenguaje de programación base, los cuales buscan explotar el hardware gráfico. En el ámbito Web, Javascript se ha convertido en unos de los lenguajes de programación dinámicos de sintaxis simple y resultados inmediatos. Por otro lado, existe una versión del API OpenGL para la Web llamada WebGL, la cual resulta una excelente opción para despliegue de gráficos 2D y 3D en la Web. Sin embargo, el proceso de inicialización y desarrollo de código en WebGL no pueden ser considerados simples.

La simplicidad de un lenguaje deriva de una sintaxis simple y sencilla. Generalmente estas características están presentes en los lenguajes interpretados, los cuales requieren un programa interprete de sus instrucciones para la ejecución de éstos. Por ello, resulta preciso contar con un lenguaje interpretado simple y conocido que explote las ventajas de los APIs gráficos existentes, y que permita reducir el esfuerzo programático para obtener salidas 2D/3D de alta calidad.

En este trabajo se presenta el desarrollo de una extensión de un lenguaje interpretado para el despliegue 3D en la Web. Dicha extensión se aplica como una herramienta que facilita el desarrollo del despliegue de gráficos, mostrando el despliegue de forma inmediata y proveyendo al usuario la posibilidad de generar una Plantilla de código en un lenguaje conocido, de manera que obtenga un despliegue básico y estándar que pueda extender o mejorar.

Nuestro aporte se puede resumir en: 1) una interfaz para la visualización de elementos 2D/3D empleando lenguaje Lua

A. Duarte, former student of Central University of Venezuela, Caracas, VE. He is now working as Senior Full-Stack Developer in Bogota, Colombia (e-mail: amarito93@gmail.com).

E. Ramirez is part of the Computer Graphics Center, Central University of Venezuela, Caracas, VE. (e-mail: esmitt.ramirez@ciens.ucv.ve).

sobre la Web; 2) un mecanismo de extracción del código de visualización realizado, en lenguaje Javascript siendo escrito en lenguaje de scripting; 3) una interfaz sencilla e interactiva para realizar prototipos de 2D/3D *rendering* de forma eficiente, en términos de líneas de código y estructura gramatical.

Este trabajo se organiza de la siguiente forma: en la Sección II se presenta una pequeña introducción al concepto de extensión de un lenguaje. La Sección III presenta los detalles de la extensión del lenguaje que se propone en este trabajo. Luego, en la Sección IV se detallan los experimentos realizados, así como los resultados obtenidos. Finalmente, la Sección IV presenta las consideraciones finales.

## II. Extensión del Lenguaje

La creación de una extensión de un lenguaje de programación conocido permite a los desarrolladores una fácil adaptación para la construcción de aplicaciones gráficas, empleando pocas líneas de código. Para ello, dicha extensión debe contar con su propio procesador del lenguaje que permita arrojar resultados visuales inmediatos. Así, es ideal que el programador pueda verificar su implementación mientras la está desarrollando (i.e. *real-time output*).

Una extensión añade nuevas características, funciones, identificadores o incluso nueva sintaxis al lenguaje original. Cuando un lenguaje define formas para extenderse a sí mismo se le conoce como lenguaje extensible. Actualmente, la mayoría de los lenguajes son extensibles en cierto grado. Por ejemplo, casi todos los lenguajes permiten al programador definir nuevos tipos de datos y nuevas operaciones (funciones o procedimientos). Algunos lenguajes permiten al programador incluir estos nuevos recursos en una unidad de programa más grande, como paquetes o módulos [1].

De acuerdo con Aho et al. [2], los lenguajes de programación son notaciones que describen los cálculos a las personas y las máquinas. Antes de ejecutar un programa, primero debe traducirse a un formato que una computadora pueda ejecutarlo; a este proceso se le conoce como procesamiento del lenguaje. Existen tres principales procesamientos, sin tomar en cuenta híbridos o combinaciones de éstos, que se pueden aplicar a los lenguajes de programación: compilación, interpretación y traducción.

La interpretación, según Sebesta [3], se encuentra en el extremo opuesto (comparado con compilación) de los tipos de procesamientos. Con este enfoque, los programas son interpretados por otro programa llamado intérprete. El intérprete actúa como un software de simulación de la máquina, su ciclo de ejecución soporta instrucciones de programas escritos en lenguajes de alto nivel en vez de instrucciones de máquina. Este software de simulación provee una máquina virtual para el lenguaje en cuestión.

En esta investigación, se optó por construir un intérprete propio de un subconjunto del lenguaje de programación Lua [4] más una extensión de este, escrito en lenguaje JavaScript [5]. Actualmente, existen diversas bibliotecas y frameworks para el despliegue de objetos 2D/3D (e.g. three.js, babylon.js, entre otras), sin embargo, estas soluciones requieren el conocimiento previo de Javascript. Nuestra solución está enfocada en programadores con conocimientos de Lua, lenguaje de script ampliamente usada en el ámbito de desarrollo de videojuegos.

En el área médica, Lua también es empleado como parte del motor de sistemas complejos en imagenología médica [6], para la interactuar con la plataforma DICOM. Como base de un sistema de visualización, Lua es empleando por MegaMol [7], un framework de prototipado para visualizaciones interactivas de grandes conjuntos de datos. Esta es escrita en C++ con el motor de *rendering* en OpenGL.

Actualmente, existen algunas propuestas basadas en framework bien conocidos en el despliegue 2D/3D como VTK [8], que liberaron ciertas funcionalidades para web (i.e. VTK.js). Para ello, es necesario conocer previamente el framework VTK y Javascript.

## III. Solución Propuesta

La solución propuesta se presenta como una herramienta web auto-contenida, es decir, no requiere hacer peticiones a un ente externo en ninguna situación. La propuesta estará formada por un intérprete de un subconjunto de Lua y un módulo denominado Desplegador de gráficos para un navegador, el cual despliega lo indicado por el intérprete. Además, permitirá generar un código en lenguaje JavaScript, con el despliegue equivalente al generado por la aplicación al interpretar el código en lenguaje Lua.

Dentro del mundo de desarrollo 3D, Lua ha surgido como un lenguaje de *scripting* bien desarrollado, flexible, de fácil uso y soportado por diversos motores gráficos. Su uso se ha extendido a aplicaciones industriales, robótica, procesamiento de imágenes, bioinformática, desarrollo web, y otros [9].

Esta herramienta, enfocada en el programador de aplicaciones gráficas, permite abrir una página web en cualquier navegador y utilizar su interfaz siguiendo el esquema que se muestra en la Fig 1.

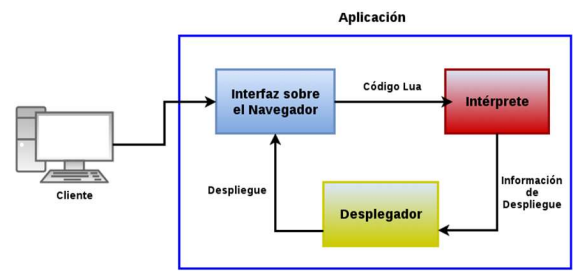

Fig. 1. Esquema general de uso de la herramienta por parte del desarrollador.

De esta forma, un desarrollador puede escribir su código en Lua empleando la extensión propuesta, la cual resulta fácil de aprender y utilizar, además de ser intuitiva. La herramienta interpreta el código y los resultados son desplegados en tiempo real. Igualmente, permite generar un archivo con el despliegue que puede ser traducido a lenguaje JavaScript (i.e. exportando el código).

La solución se dividió en tres fases funcionales: procesamiento de lenguaje, despliegue gráfico e interfaz de usuario.

5606

## A. Procesamiento del Lenguaje

El procesamiento del lenguaje consta de diversos procesos que permiten reconocer un lenguaje. Estos se pueden describir de forma secuencial: diseño de la gramática, analizador léxico, analizador sintáctico y analizador semántico.

Para el diseño de la gramática se optó por un lenguaje de scripting interpretado para agilizar su ejecución a medida que el usuario agrega instrucciones al código. A la gramática de Lua se le agregó una extensión para el despliegue de gráficos, siendo un subconjunto lo suficientemente amplio de la gramática. Esto resulta suficiente para hacer uso de la extensión propuesta.

La extensión para el despliegue de gráficos permite desplegar objetos 2D/3D predefinidos, aplicar transformaciones afines básicas, cargar modelos en formato OBJ, aplicar modelos de sombrado Blinn-Phong, Gouraud y Flat, así como los componentes de la iluminación: ambiental, difuso y especular. La gramática utilizada en notación BNF (Backus–Naur form) se muestra en la Fig. 2.

```
1  start ::= chunk
2  chunk ::= {stat}[last_stat]
3  last_stat ::= 'return' [exp_list][';']
4  stat ::= 'local' name_list ['=' exp_list] |';' |var_list '=' exp_list |func_call |
5         'while' exp 'do' chunk {'elseif' exp 'then' chunk} ['else' chunk] |
6         'for' name '=' exp ',' exp [',' exp] 'do' chunk 'end' |'repeat' chunk 'until' exp |
7         'function' func_name func_body | 'break' | line_comment | long_comment
8  func_name ::= name
9  var_list ::= var {',' var}
10 var ::= name {('.' name | '[' exp ']')}
11 name_list ::= name {',' name}
12 exp_list ::= exp {',' exp}
13 exp ::= exp bin_op exp | un_op exp | '(' exp ')' |'nil' |'true' |'false' |
14       string |char_string |numeral | var | func_call | hash_const
15 bin_op ::= 'or' | 'and' | '==' | '~=' | '>=' | '>' | '<' | '|' |
16         '~' | '&' | '>>' | '<<' | '..' | '-' | '+' | '%' | '//' | '/' | '*' | '^'
17 un_op ::= '~' | '-' | '#' | 'not'
18 func_call ::= var args
19 args ::= '(' [exp_list] ')'
20 hash_const ::= '{' [field_list] '}'
21 field_list ::= field {field_sep field}
22 field_sep ::= ';' | ','
23 field ::= '[' exp ']' '=' exp | name '=' exp | exp
24 func_body ::= '(' [name_list] ')' chunk 'end'
```

Fig. 2. Gramática empleada en nuestra propuesta basada en Lua.

Igualmente, para el despliegue de los gráficos se emplea la extensión propuesta en la Fig. 3.

```
1   DrawCube (String displayMode);
2   DrawCone (String displayMode);
3   DrawSphere (String displayMode);
4   DrawCylinder (String displayMode);
5   DrawGrid (String displayMode);
6   DrawObject (String displayMode);
7   TranslateObject (Array vector);
8   RotateObject (Number angle, Array vector);
9   ScaleObject (Array vector);
10  DrawPointLight (Array pos);
11  DrawDirectionalLight (Array pos, Array dir);
12  DrawSpotLight (Array pos, Array dir, Number cutoff, Number expor
13  ChangeLighting (String model);
14  AmbientalComponent (Array vector);
15  DiffuseComponent (Array vector);
16  SpecularComponent (Array vector);
```

Fig. 3. Extensión propuesta para el despliegue de gráficos.

Así, tomando como ejemplo la instrucción *TranslateObject* (línea 7), ésta representa a una función que mueve un objeto desde su posición actual, hasta lo indicado por el vector del tipo Array (parámetro formal).

El análisis léxico se realizó con la herramienta Jison [10], la cual permite generar un código en lenguaje Javascript equivalente al parser que emplea la gramática diseñada. El analizador léxico identifica los lexemas que se utilizan en el lenguaje.

El análisis sintáctico fue realizado con la misma herramienta Jison. En comparación con el analizador léxico, este se compone por reglas gramaticales, las cuales definen la sintaxis que debe cumplir todo código evaluado por el parser. Igualmente, utiliza los lexemas identificados por el analizador léxico en sus reglas para identificar cuál regla gramatical equivale a la instrucción evaluada.

Jison provee una sintaxis parecida a BNF para escribir la gramática que utilizará el parser, pero a diferencia de BNF en esta sintaxis se requiere establecer un orden claro de evaluación de las reglas. Para llenar las estructuras que empleará el analizador sintáctico, Jison provee bloques de código que pueden colocarse al final de la regla gramatical. De esta manera, se ejecutan en el lenguaje destino una vez que se identifique la regla gramatical.

El analizador a medida, que identifica instrucciones válidas, genera nodos de una estructura llamada AST o Árbol de sintaxis abstracta. Se utilizó un analizador sintáctico ascendente para la generación de esta estructura. Esta estructura fue utilizada en siguiente analizador semántico quien se encarga de llenarla de datos. En la Fig. 4 se ilustra un ejemplo del AST.

Por último, el análisis semántico recorre el AST generado, construyendo una tabla de símbolos, administrando los ambientes de ejecución o alcance de las variables, invocando al Desplegador gráfico, y ejecutando las instrucciones en lenguaje Lua para obtener los parámetros que el Desplegador empleará.

El parser es un código en lenguaje JavaScript, el cual se recibe como parámetro el código del usuario por la aplicación en su invocación. Una vez ejecutados los análisis del parser, la aplicación invoca a una función llamada eval, que se encarga de recorrer recursivamente el AST en profundidad, es decir, los últimos nodos u hojas son los evaluados primero.

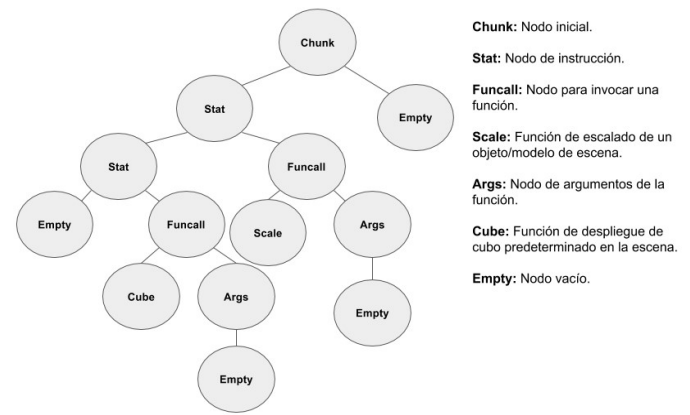

Fig. 4. Ejemplo de un Árbol de Sintaxis Abstracta empleada por Jison.

Cada instrucción dentro del lenguaje Lua simboliza un tipo de nodo que puede encontrarse en el AST, y por tanto la estructura contiene instancias de estos. Nótese que la diferencia entre dos nodos del mismo tipo radica en la información que contengan para ser ejecutadas.

### B. Despliegue Gráfico

En el despliegue gráfico se considera una escena como el entorno donde se describen todos los entes visuales que se mostrarán al usuario. Al considerar el uso de un navegador, el Desplegador emplea el API WebGL [11] para el despliegue de gráficos. Con este API es posible inicializar los parámetros de una escena: cámara, modelos y transformaciones. Cabe destacar que se emplea WebGL sin el uso de frameworks o bibliotecas externas para ofrecer Plantillas de código sin convenciones, i.e. plantillas base para que el usuario tenga el control de todos los procesos del despliegue.

Dentro de la escena, existe una cámara ubicada en una posición en $R^3$ asociado a una dirección y rango (amplitud) de visión. La cámara puede moverse de forma libre dentro de la escena. En la escena existen objetos (representados como modelos geométricos de vértices y aristas) que son desplegados empleando programas para la tarjeta gráfica (i.e. 3D shaders).

El Desplegador cuenta con la implementación del modelo de iluminación Phong [12] para los objetos dentro de la escena. De este modo, existen característica de los materiales de los objetos, así como ciertas propiedades de las fuentes de luz. El usuario es libre de agregar diversas luces, objetos, transformaciones, entre otros.

Los modelos de sombreado de polígonos Flat, Gouraud y Blinn-Phong son soportados por el Desplegador; así como tres tipos de comportamiento de las fuentes de luz: puntual, direccional y concentrada [12].

Por su parte, un modelo geométrico está compuesto, además de vértices y aristas, por vectores normales por cada vértice. Igualmente, se requiere el manejo de matrices para la transformación de un espacio a otro (e.g. de modelo, normal, etc.), y funciones para la inicialización de los búferes a ser enviados a los shaders.

Nuestra propuesta posee de forma predefinida cinco (5) modelos: cubo, cono, esfera, cilindro y grid. Estos modelos pueden ser desplegados inmediatamente, sin necesidad de una configuración inicial. También, la aplicación permite la carga de modelos en formato OBJ (Object Format File). Estos modelos se cargan con un módulo diseñado ad-hoc. Las normales por vértice son aproximadas mediante el promedio de las normales por cara.

Los modelos poseen una propiedad de despliegue que el usuario puede modificar, la cual realiza su despliegue empleando triángulos, puntos o líneas. Así, los usuarios pueden visualizar la malla del modelo. Adicionalmente, todo modelo posee un material que da información a los shaders sobre las características del color del modelo.

### C. Interfaz de Usuario

La interfaz de usuario está construida con el objetivo de proveer al programador un editor de código, una ventana para visualizar los gráficos, una consola de solo lectura para los mensajes de la solución, y un menú con opciones. La Fig. 5 ilustra cada uno de los componentes presentes en la interfaz, como lo son la ventana para visualizar código, el editor, una barra de menú y la consola de salida para el usuario.

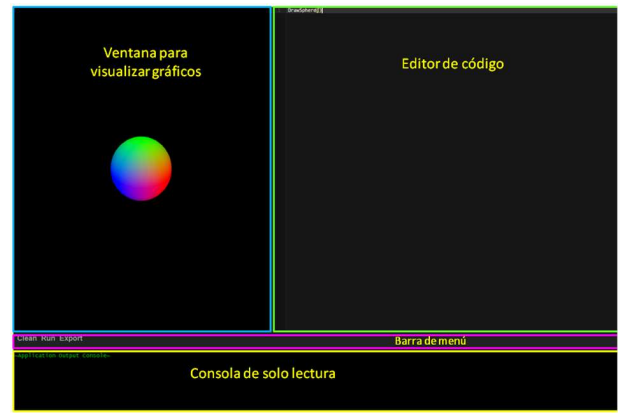

Fig. 5. Interfaz gráfica del editor propuesto en un navegador.

La interfaz está basada en HTML5, CSS3 y JavaScript. La interfaz se basa en la técnica *responsive*, la cual ajusta los componentes de la interfaz a medida que la ventana del navegador cambia de tamaño, ajustándose a la pantalla de casi cualquier tipo de dispositivo de salida.

Para escribir código utilizando la extensión propuesta, se proporciona un editor que muestra las líneas de código enumeradas, enmarca en color rojo la línea que contenga un error sintáctico, y permite el uso de los accesos directos como copiado y pegado.

Se diseñó una estructura de código llamada Plantilla que contiene el código que genera el despliegue gráfico desarrollado por el usuario, sin componentes de interfaz ni módulos de la aplicación que no se relacionen directamente con el despliegue de gráficos. Así, el usuario puede obtener un código para su uso sin ningún obstáculo como módulos innecesarios, código ofuscado o desorganizado. El usuario es capaz de mejorar o modificar a su voluntad el despliegue desarrollado rápidamente utilizando la aplicación propuesta.

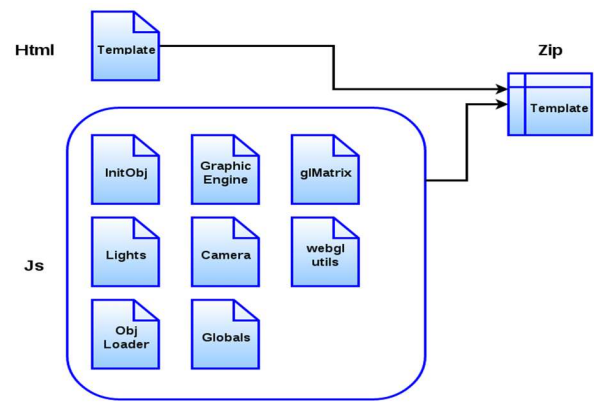

Fig. 6. Estructura de una Plantilla, compuesta por los archivos Javascript y un archivo base HTML.

El usuario tendrá disponible estas Plantillas en forma de un archivo comprimido, conteniendo cada módulo de despliegue de gráficos por separado y un archivo con la información de cada objeto/modelo desplegado en la escena. Cada uno de estos módulos es una Plantilla, es decir, un código en lenguaje JavaScript utilizando el API WebGL organizado de manera que sea comprendido por el usuario, y con nombres mnemónico de variables y funciones. En la Fig. 6 se muestra la estructura que presenta una Plantilla.

Es de importancia entender que no se ejecuta una traducción del código escrito en lenguaje Lua a lenguaje JavaScript, sino del despliegue desarrollado en la aplicación a un despliegue hecho en JavaScript con el API WebGL utilizando sólo lo necesario para su ejecución. Esto añade eficiencia en cuanto a que el resultado esperado sea el óptimo.

Por ejemplo, en la Fig. 7 se muestra un fragmento de código escrito en la extensión de Lua dentro del editor, donde se observan cinco objetos desplegados, y cuatro de ellos trasladados con la función TranslateObject.

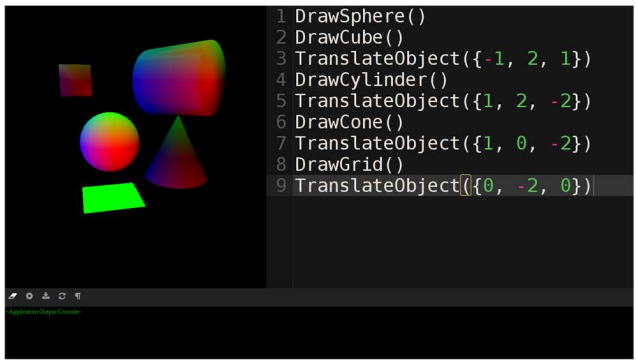

Fig. 7. Ejemplo de cinco objetos desplegados aplicando transformaciones.

## IV. PRUEBAS Y RESULTADOS

La experimentación realizada busca medir el esfuerzo programático para generar una salida gráfica. Igualmente, se construyen pruebas de rendimiento y comparación entre la salida obtenida y la generada por las plantillas.

La solución propuesta permite desplegar figuras en 2D y 3D (predeterminadas o en formato OBJ), los tres modelos de sombreado mencionados, cambiar el componente ambiental, difuso y especular tanto en modelos como luces, emplear cambios de colores bajo un esquema RGB, y las transformaciones afines rotación, escalamiento y traslación. Para todas estas acciones se incluyeron once (11) funciones con parámetros en la sintaxis de Lua, necesitando solo el 39.32% del total de reglas gramaticales de la sintaxis de Lua.

Por otro lado, el código se encuentra disponible de forma pública en un repositorio de Github [13], para su descarga y mejora continua

### A. Esfuerzo Programático

Se diseñaron distintos casos de estudio donde se mide la cantidad de líneas de código empleando nuestra propuesta, y empleando código en WebGL de forma regular. Dado que existen diversas maneras de escribir código en WebGL, se crearon tres códigos distintos, de tres programadores distintos, para promediar el número de líneas de código.

Para desplegar un objeto predeterminado (ver Sección III-B), se requiere en promedio 415 líneas de código entre JavaScript y WebGL. Con nuestra propuesta, se requieren menos de 5 líneas. Es claro que, al ser una biblioteca con diversas funciones ya escritas, la cantidad de código es extremadamente menor.

También, se realizaron pruebas para el despliegue de objetos en formato OBJ aplicando luces/transformaciones, y desplegando tres objetos y tres luces. Esta última, requirió aproximadamente 1015 líneas de código de manera convencional versus 15 líneas de nuestra propuesta.

### B. Pruebas de rendimiento

Una primera prueba consistió en desplegar 10, 100 y 1000 esferas de 2880 triángulos, en tres navegadores distintos: Google Chrome, Mozilla Firefox y Opera. computadora máquina de pruebas empleó un procesador Intel i5-3210M, una memoria RAM de 16 Gb bajo el sistema operativo Debian.

La tabla I presenta los resultados obtenidos medidos en ms, donde debido al bajo valor de los tiempos se decide mostrar su valor mínimo y máximo. Entre Chrome y Firefox, los tiempos son muy similares, y Opera ligeramente requiere mayor tiempo. Sin embargo, eso no afecta directamente al rendimiento como para concluir un impacto considerable por el navegador.

TABLA I
VALORES MÍNIMOS Y MÁXIMOS EN TRES NAVEGADORES PARA 10, 100, Y 1000 ESFERAS.

|  | Chrome | | Firefox | | Opera | |
| --- | --- | --- | --- | --- | --- | --- |
|  | mín | máx | mín | máx | mín | máx |
| 10 | 1 | 2 | 1 | 5 | 1 | 7 |
| 100 | 1 | 5 | 1 | 6 | 1 | 9 |
| 1000 | 4 | 26 | 7 | 17 | 8 | 34 |

Similarmente, debido al valor bajo de los tiempos y a la cantidad de muestras se decidió tomar el mínimo y máximo del tiempo. Se puede observar que entre los dos (2) primeros casos de prueba hay una diferencia baja de tiempo en cualquier navegador, y el último caso presenta un mayor tiempo. También se puede observar que el navegador Opera presenta un mayor tiempo en todos los casos de prueba.

Sin embargo, para establecer una comparación al utilizar nuestra solución, se consideró el tiempo desde que se ejecuta el código hasta que se genera la salida gráfica correspondiente. Luego, se comparan los tiempos y se ilustra la diferencia de tiempo entre ejecutar un despliegue y ejecutar un despliegue luego de hacer un procesamiento de un lenguaje.

La tabla II muestra los promedios obtenidos de ejecutar cada caso de prueba con procesamiento de lenguaje tomando cinco muestras de tiempo. Nótese que la mayor cantidad de tiempo se obtiene con el navegador Firefox.

TABLA II
TIEMPO PROMEDIO DESDE EL PROCESAMIENTO HASTA LA VISUALIZACIÓN.

|  | Chrome | Firefox | Opera |
| --- | --- | --- | --- |
| 10 | 72.8 ms | 69.6 ms | 74.0 ms |
| 100 | 311.8 ms | 520.8 ms | 315.8 ms |
| 1000 | 2169.0 ms | 4334.6 ms | 2210.8 ms |

5606

La Fig. 8 ilustra la diferencia de tiempo entre los tiempos de despliegue y los tiempos desde el procesamiento del código hasta el despliegue. Se aprecia la diferencia de tiempo que añade el procesamiento del código en los casos de prueba. Sin embargo, este tiempo añadido sólo se presenta al interpretar el código. Una vez que el código haya sido interpretado, todos los *frames* a continuación consumirán sólo tiempo de despliegue, pues sólo se interpreta una vez el código y se almacena la información de la escena para ser desplegada periódicamente.

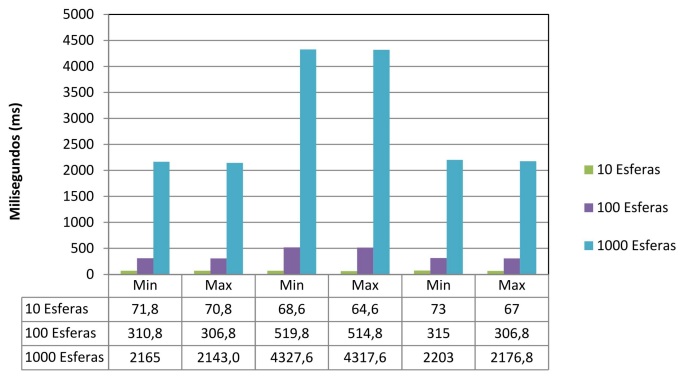

Fig. 8. Resultados de las pruebas realizadas por 10, 100 y 1000 esferas.

Por otro lado, parte de las pruebas evaluaron la cantidad de recursos consumidos (CPU y RAM), desde momento de procesamiento del código en la extensión propuesta hasta la ejecución del primer *frame* del despliegue. La Fig. 9 muestra los recursos consumidos.

El porcentaje de recursos consumidos por la aplicación entre los dos primeros casos de prueba no presentan una gran diferencia. Sin embargo, el último caso de prueba presenta un gran consumo de recursos tanto la CPU como memoria. Además, el navegador Firefox presenta un mayor consumo de recursos en todos los casos de prueba.

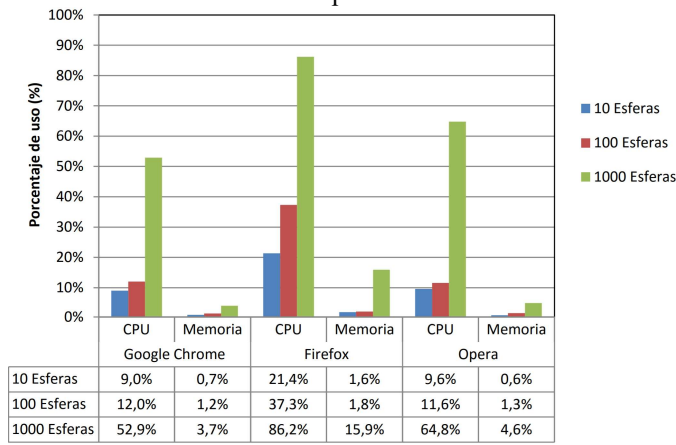

Fig. 9. Resultados del consumo de recursos en tres distintos navegadores.

### A. Comparación

La solución propuesta genera una plantilla, es decir, un código en Javascript que genera la misma salida que se está despliega en el editor. Entonces, se realizaron comparaciones entre la salida visual en el editor y la salida visual que se obtiene de una plantilla en un navegador.

Aplicando transformaciones e iluminación sobre objetos predeterminados y cargados desde archivos OBJ, la salida visual es la misma. Para determinar este factor, se empleó la herramienta *Perceptual Image Diff* (pdiff.sourceforge.net) la cual permite comparar dos imágenes bajo el sistema visual humano. Esto demuestra que nuestra propuesta arroja los resultados de forma robusta y consistente.

### V. CONCLUSIONES.

En este trabajo se presentó una extensión para el lenguaje interpretado Lua con el fin de desplegar objetos 3D en un browser. Esto permitió reducir drásticamente el esfuerzo programático debido a la disminución en el número de líneas necesarias para obtener una salida gráfica. Nuestra extensión facilita el desarrollo de escenas complejas, debido a que provee el uso de estructuras iterativas y de control necesarias en la programación estructurada.

Del mismo modo, se implementaron funcionalidades basadas en un subconjunto del lenguaje Lua como lo demuestran las pruebas realizadas. Esto permitió añadir efectos básicos en el *rendering* de objetos 3D que son de suma utilidad en el despliegue de objetos. Consideramos que a pesar de que Lua sea un lenguaje de scripting sencillo, podría ocasionar resistencia de cambio con algunos programadores de WebGL.

Basados en nuestras pruebas podemos analizar que la diferencia de tiempo añadido debido al procesamiento del código es aceptable, ya que el tiempo de espera por una salida gráfica sólo se ve afectado al desplegar más de 3000 triángulos. Entonces, considerar modelos más complejos pueden verse afectados. Sin embargo, emplear un hardware de mayor procesamiento y rendimiento, podría implicar aumentar el límite propuesto.

Por otro lado, gracias a las Plantillas generadas, se pueden estudiar y utilizar este Código para generar proyectos de software libre que promuevan la evolución de las técnicas para el despliegue de gráficos, o ser empleadas junto con otras bibliotecas [14]. Las Plantillas generadas despliegan los mismos objetos/modelos con las mismas características que se desarrollaron, garantizando así una equivalencia entre la salida gráfica generada por la aplicación y por las Plantillas.

En un futuro, se propone extender el uso de la extensión al añadir más funcionalidades propias del realismo en un ambiente 3D. . Adicionalmente, se propone construir un editor con una mayor cantidad de opciones (e.g. control de usuarios, librería de objetos, etc.), tal como el que se presenta en [15]. Creemos que de esta manera se podrá expandir y aumentar su utilización.

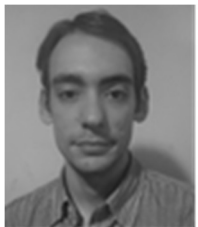
**Amaro Duarte** born in Caracas, Venezuela on the 27th March, 1993. He received the Bachelor degree in Computer Science in 2016 from Universidad Central de Venezuela - UCV, Caracas, Venezuela. At the present, he is working as Senior Full-Stack Developer at Hexacta in Bogotá, Colombia. His mainly projects are related to Node.js technology.

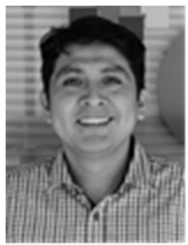
**Esmitt Ramírez** born in Lima, Perú on the 27th October 1981. He received both the Bachelor degree in Computer Science in 2005, and the Magister Scientiarum in 2011 from Universidad Central de Venezuela - UCV, Caracas, Venezuela. Currently is member of the Computer Graphics Center, UCV as lecturer, and PhD student in the Computer Vision Center in the Autonomous University of Barcelona, Spain. His mainly researches are focused on digital image processing, medical imaging and *rendering* algorithms.